\documentclass[runningheads]{llncs}
\usepackage[utf8]{inputenc}
\usepackage{float}
\usepackage{array}
\usepackage{multirow}
\usepackage{graphicx}
\usepackage{amsmath}
\usepackage{color, colortbl}

\definecolor{Gray}{gray}{0.90}

\newcommand\MyBox[1]{
	\fbox{\lower0.75cm
		\vbox to 1.7cm{\vfil
			\hbox to 1.7cm{\hfil\parbox{1.4cm}{\centering \large#1}\hfil}
			\vfil}%
	}%
}

\begin{document}

\title{Detecting Fraudulent Accounts on Blockchain: \\ A Supervised Approach}

\titlerunning{Abbreviated paper title}
% If the paper title is too long for the running head, you can set
% an abbreviated paper title here
%
\author{Michał Ostapowicz \and
Kamil Żbikowski}

\institute{Institute of Computer Science\\Faculty of Electronics and Information Technology\\Warsaw University of Technology\\
ul. Nowowiejska 15/19, 00-665 Warsaw, Poland\\
\email{michal.ostapowicz.stud@pw.edu.pl}
\email{kamil.zbikowski@ii.pw.edu.pl}
}

%
% First names are abbreviated in the running head.
% If there are more than two authors, 'et al.' is used.
%
\maketitle              % typeset the header of the contribution

\begin{abstract}
Applications of blockchain technologies got a lot of attention in recent years. 
They exceed beyond exchanging value and being a substitute for fiat money and traditional banking system. 
Nevertheless, being able to exchange value on a blockchain is at the core of the entire system and has to be reliable. 
Blockchains have built-in mechanisms that guarantee whole system's consistency and reliability. 
However, malicious actors can still try to steal money by applying well known techniques like malware software or fake emails. 
In this paper we apply supervised learning techniques to detect fraudulent accounts on Ethereum blockchain. 
We compare capabilities of Random Forests, Support Vector Machines and XGBoost classifiers to identify such accounts basing on a dataset of more than 300 thousands accounts. 
Results show that we are able to achieve recall and precision values allowing for the designed system to be applicable as an anti-fraud rule for digital wallets or currency exchanges. 
We also present sensitivity analysis to show how presented models depend on particular feature and how lack of some of them will affect the overall system performance.
\end{abstract}
\section{Introduction}
Recent developments in digital currencies gave birth not only to a completely new way of exchanging value, but also to such areas like distributed trust management. Those advances may replace traditional notary services or payment processing companies in the near future \cite{worner2016bitcoin}.  Such advances are possible to achieve thanks to technology called blockchain that, in its basis, is as an immutable, distributed database.  First public blockchain, called Bitcoin, was launched in 2009 and, not surprisingly, from the very beginning attracted fraudulent actors that tried to take advantage of other participants. These actors very often try to convince others to send them digital currency to their accounts by using different techniques like malware or fake emails. Due to the publicly available data, information about account once denoted as fraudulent can be shared and available without limitations. Quite contrary to traditional financial systems, all the transfers to and from such account can be freely viewed and analyzed. The availability of this data gives us an opportunity to verify if there is a meaningful relation between operations done on the account and this account being fraudulent.

In this paper, we propose a novel approach for detecting fraudulent accounts on Ethereum network. Ethereum is a blockchain that has some significant improvements over Bitcoin \cite{buterin2014next}. Those improvements allow to write and execute contracts (called smart contracts) more easily. These contracts give an opportunity for many different actors to engage in complex agreements that are fully executable and can be verified with the use of the underlying protocol. More details on Ethereum can be found in \cite{wood2014ethereum}. 

In the first stage, we automatically gathered available data about accounts and transactions. Then, we created explanatory variables out of raw data. They represent aggregates and statistics computed over volumes and time. In the next stage, we tested three classifiers and compared their results in the context of possible applications. They can strongly depend on different use cases that may put more importance on precision than on recall or the other way round. The contribution of this study can be summarized as follows:
\begin{itemize}
    \item We proposed a novel approach for identifying fraudulent accounts on Etherum blockchain that is easily transferable to other blockchains, like Bitcoin. 
    \item  We  conducted a thorough analysis of three different machine learning algorithms for the task of classification accounts to “fraudulent” or “not fraudulent” class.
    \item  We conducted a sensitivity analysis in order to verify how much we depend on particular explanatory variables. This is a test that allow us to address the potential problem of a look-ahead bias that may or may not exist within the data that we gathered. 
\end{itemize}

\section{Related work}
\label{sec:rel}
Detecting fraudulent activity in financial operations is a well known problem. Both researchers and practitioners put a lot of attention to developing new tools that would correctly identify new attack vectors. This is an endless battle in which both sides use their creativity and new technologies. A comprehensive survey on fraud detection techniques can be found in \cite{kou2004survey}. More recent surveys on fraud prevention systems and detecting financial fraud through data mining algorithms can be found in \cite{abdallah2016fraud} and \cite{bhardwaj2016financial} respectively. 

Quah and Sriganesh \cite{QUAH20081721} used Self Organizing Maps (SOM) to detect credit card frauds. They took an approach that if a transaction is similar to all transactions in a set of genuine transactions, it is also considered genuine. On the other hand, if it looks like any of the transactions in a set of fraudulent, then it is also considered fraudulent. In addition to the basic task of clustering input data, Self Organizing Maps are also used to detect and extract hidden patterns. According to the authors, in real financial systems that verify each transaction on multiple layers, SOM may also serve as a filter for the layers following it. In the case described by the authors, SOM receives an input data vector consisting of client, account and transaction features.

In \cite{CARNEIRO201791} authors used supervised learning methods to tackle similar problem.
They used logistic regression, Support Vector Machine (SVM) and random forest.
Apart from using typical transaction features as an algorithm's input (e.g. order value, type of items ordered, payment method), through abstraction and combination they engineered several new variables such as binary evaluated compliance of the country of the card transaction with the country to which the purchased items are to be delivered.
Eventually, the authors used 71 features to describe each transaction. 
The best results were obtained using random forest method, which is why it was used in further analysis. 
As it turned out, despite quite good results in recognizing frauds, they were not good enough to fully automate verification of transactions.

In case of transfers done through blockchain transactions, fraud detection can be a more complicated task as most of the time we are not in possession of geographical and personal data of participants. Pham and Lee \cite{DBLP:journals/corr/PhamL16} in their article dealt with detecting frauds in the Bitcoin network.
The network data was modeled as two graphs: a user graph and a transaction graph which were used to detect anomalies (e.g. fraudulent and suspicious users). 
They had information about 30 cases of theft in the Bitcoin network, which were later used to verify their results. 
In both graphs, each vertex was represented with 12 features, such as the input and output stage, the average time between transactions, the creation date and activity time. 
As the first step in the analysis they applied k-means algorithm to group all graph nodes. 
As the authors pointed out, this algorithm is not used to find anomalies, but it may be useful, because the points that diverge from the rest are expected to be found far from the centroids calculated with k-means algorithm. 
They wanted to investigate if anomalies in user graph, clearly refer to anomalies in the transaction graph, i.e. whether "suspicious" users were involved in "suspicious" transactions.
To find anomalies in these groups authors used a method based on the Mahalanobis distance and Support Vector Machine (SVM). 
Suspected users and transactions indicated by both algorithms overlapped to a large degree.
In both methods extreme values were indicated as suspicious, i.e. vertices with the largest or smallest degrees. 
That approach allowed to detect two authentic anomalies: one theft (detected by the Mahalanobis distance based method) and one loss caused by a corruption in a hashing function (detected by the SVM).
These results do not seem to be statistically significant primarily due to a limited number of known thefts (or anomalies in general). 
\section{Methodology}
\subsection{Data preparation}
The data used in the analysis came from the Etherscan.io website, which is one of the most popular Ethereum blockchain browsers.
It provides information about all transactions in the network, mined blocks and user accounts. 
Over 2\,500 wallets were reported by the users as related to illegal activities and marked as "Hack/Phishing". 
Using the Etherscan API it was possible to download information about all transactions in which given wallet participated. 
Some of the wallets tagged as fraudulent had no transactions at all or were involved mostly in ERC20 token trade. They were not included in the dataset. After this correction we analyzed 2\,200 wallets marked as involved in illegal activity.
In addition to fraudulent transactions data, we also collected information about transactions from 349\,999 randomly selected wallets out of the 65\,564\,460 existing (as of 28th May 2019) in the Ethereum network. They were not marked as suspicious and were considered non-fraudulent.

Based on the work of \cite{DBLP:journals/corr/PhamL16} we decided to create 13 explanatory variables concerning transaction data of each account. Explanatory variables are presented in table \ref{tab:vars}.
\begin{table}%%
	\caption{Explanatory variables}
	\label{tab:vars}
	\centering
	\begin{tabular}{ll}
		\hline
		Variable name&Variable description\\
		\hline
		IT&amount of incoming transactions\\
		OT&amount of outgoing transactions\\
		UIT&amount of unique incoming transactions\\
		UOT& amount of unique outgoing transactions\\
		AVIT&average value of the incoming transaction\\
		AVOT&average value of the outgoing transaction\\
		VIT&total value of all incoming transactions\\
		VOT&total value of all outgoing transactions\\
		ATIT&average time between incoming transactions\\
		ATOT&average time between outgoing transactions\\
		AGP&average gas price\\
		AGL&average gas limit\\
		DUR&active duration (time in days since the first until the last transaction)\\
		\hline

	\end{tabular}
	\vspace{-1mm}   
\end{table}
The dataset was divided into two parts: a training set with 281\,760 samples and a validation set with 70\,439 samples. 

\subsection{Experiment setup}
\begin{figure}%[H]
	\includegraphics[width=\linewidth]{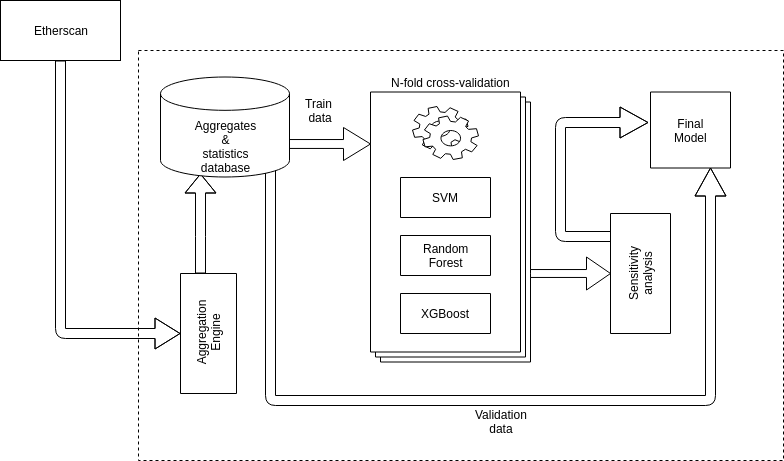}
	\caption{System and data architecture for the conducted experiment}
	\label{fig:process}
\end{figure}

The prediction problem definition here is a classic example of a binary classification. We examined following classifiers: Random Forests, Support Vector Machines and XGBoost in order to determine their capabilities of making accurate predictions for a given dataset. Figure \ref{fig:process} presents data and system architecture for the conducted experiment. As a first step we downloaded data using the Etherscan API, which then was aggregated to create 13 variables presented in the Table \ref{tab:vars}. In the next step, using grid search with 10-fold cross-validation we tried to find set of parameters that could give the best results for the three supervised learning algorithms that we chose.

Data gathered from Etherscan did not allow to accurately determine the moment of marking particular account as a fraudulent one. It can be possible that certain aggregates that we use for training are biased and data used to compute them was gathered after the moment of marking a particular account as fraudulent. It is possible that some of the transactions can be a result of the public exposure of an account. This would not be a problem if were only interested in devising a method for simple classification of account. However, if we would like to use proposed method as an early warning system then we will have to take a moment of an exposure into consideration. We address this issue by conducting performance analysis after removing most important explanatory variables. As the final step we did a validation check on a part of a dataset that was not used for the training purposes. Result from this step were reported in the following sections.

\subsection{Prediction models}

%Dodać źródła
\par The Support Vector Machine (SVM) classifier is a binary classifier algorithm that looks for an optimal
hyperplane as a decision function in a high-dimensional space \cite{boser1992training}. Having a training dataset $\{\mathbf{x}_k,y_k\} \in \rm I\!R^n \times \{-1,1\}$
where $\mathbf{x}_k$ are the training examples and $y_k$ are the class labels at first we map $\mathbf{x}$
into a higher dimensional space via a function $\Phi$, then computing a decision function in the form of:
\begin{equation}
f(x) = \langle \mathbf{w},\Phi(\mathbf{x})\rangle+b
\end{equation}
by maximizing the distance between the set of points $\Phi(\mathbf{x}_k)$ to the hyperplane parameterized by
$(\mathbf{w}, b)$. The class label of $\mathbf{x}$ is given by the
sign of $f(\mathbf{x})$. The optimization problem for the SVM classifier with penalized misclassified examples can be written as:
\begin{equation}
\min_{\mathbf{w} , \xi} \frac{1}{2} ||\mathbf{w}||^2 \\ + \sum_{i=1}^{m} C\xi_i,
\label{eq:ww}
\end{equation}
subject to:
\begin{equation}
y_i f(\mathbf{x}_i) \geq 1-\xi_i,
\end{equation}
With variables $\alpha_i$ defined such that:
\begin{equation}
\mathbf{w} = \sum_{i}^{m} \alpha_i y_i \mathbf{x_i},
\end{equation}
by solving for the Lagrangian dual of the problem \ref{eq:ww}, we obtain the simplified problem:
\begin{equation}
\max_\alpha Q(\alpha) = \sum_{i=1}^{m} \alpha_i - \frac{1}{2}\sum_{i=1}^{m}\sum_{j=1}^{m} \alpha_i\alpha_jy_iy_j\varphi(\mathbf{x_i})\varphi(\mathbf{x_j})
\end{equation}
subject to:
\begin{equation}
\sum_{i=1}^{m} \alpha_i y_i = 0,
\end{equation}
\begin{equation}
\alpha_i \geq 0.
\end{equation}
\par Random Forest is a classifier consisting of a collection of tree-structured
classifiers $\{h(\mathbf{x}, \Theta_k), k = 1, ...\}$  where $\{\Theta_k\}$  the are independent identically distributed
random vectors and each tree casts a unit vote for the most popular class at input. \cite{breiman2001random}
For each tree in the random forest new training set is generated, by drawing with replacement from the original training set.
Tree is grown on the new training set using random feature selection at each node. The resulting trees are not pruned.
\par XGBoost is a scalable machine learning system for tree boosting proposed by Chen and Guestrin \cite{DBLP:journals/corr/ChenG16}. 
The impact of this system has been lately recognized in a number of machine learning and data mining challenges. For example, among the 29 challenge winning solutions published at Kaggle’s blog during 2015, 17 solutions used XGBoost. 

Considering training dataset $\{\mathbf{x}_k,y_k\} \in \rm I\!R^n \times \rm \{-1,1\}$
where $\mathbf{x}_k$ are the training examples, $y_k$ are the class labels and $n$ is number of features, the output of model is voted or averaged by a collection $F$ of $k$ regression trees:
\begin{equation}
\hat{y}_i = \phi(\mathbf{x}_i) = \sum_{i=1}^{k}f_k(\mathbf{x}_i), f_k \in F
\end{equation}
Each regression tree contains a continuous score on each of the leaves ($w_i$ represents score on the $i$-th leaf).
To learn the set of functions used in the model, the following objective needs to be minimized
\begin{equation}
L(\phi) = \sum_{i}l(\hat{y}_i, y_i) + \sum_{k}\Omega(f_k) \\
\end{equation}
$l$ is the training loss function which measures how well the model fits on training data. The second term $\Omega$ penalizes the complexity of the model and is defined as:
\begin{equation}
\Omega(f) = \gamma T + \frac{1}{2} \lambda ||w||^2 \centering
\end{equation}
where the $\gamma$ is the complexity of each leaf, $T$ is the number of leaves in a decision tree and $\lambda$ is a parameter to scale the penalty. If we apply the second-order Taylor expansion to the loss function and remove the constant terms we obtain the objective at the $t$-th iteration in the form of:
\begin{equation}
\label{eq:xgb}
\tilde{L}^{(t)} = \sum_{i=1}^{n}[g_i f_t(\mathbf{x}_i)+\frac{1}{2}h_i f^2_t(\mathbf{x}_i)] + \Omega(f_t)
\end{equation}
where $g_i$ and $h_i$ are respectively first and second derivative of the loss function.

\section{Empirical results}
Our objective was to find a prediction model that could be used as a real-world fraud detection system. Due to the high class imbalance we decided to focus our assessment of a particular algorithm on analyzing recall and precision statistics. For different parameters configurations we obtained results with either high recall and low precision or low recall and high precision. The former one has an obvious advantage of capturing most of the frauds that were present in a dataset. On the other hand, it is completely useless for a real world applications in which all the alerts have to be manually analyzed by a human being. 

As we included almost all of the fraudulent transaction and only minor sample of non-fraudulent, we had distribution in which probability of a random account being a fraudulent one was significantly higher than in the real-world. Because of that, we could not rely on precision statistic as it is vulnerable to this problem. Instead of using precision as a false alarm verification cost estimator we decided to use false positive rate. It fits our purpose since it does not depend on the total amount of frauds in the dataset.

\subsection{Random forest results}
For random forest we decided to tune number of variables randomly sampled as candidates at each split (mtry), minimum size of terminal nodes (min.node.size) and different cut-off probabilities i.e. probability above which sample is actually predicted as a non-fraud.

As we can see in Table \ref{tab:rf_val_results} biggest impact on the results has threshold which determines final predicted class. Larger threshold causes less samples to be classified as non-fraud and therefore an increase of recall and at the same time increase in FPR which we would like to keep low.

Instead of choosing one configuration which would be a trade-off between recall and false positive rate, we decided to distinguish classifiers able to find as many actual fraudulent accounts as possible (maximizing recall) and a classifiers that make as few mistakes in predicting fraud class as possible (minimizing false positive rate). 
Validation results presented in Table \ref{tab:rf_val_results}, are similar to the ones we got with cross-validation and confirm, the best configurations are: Conf. 3 in terms of FPR and Conf. 19 in terms of recall. For chosen configurations of random forest we created confusion matrices (presented in tables \ref{tab:rf_conf1} and \ref{tab:rf_conf2}) that help to better analyze performance of this classifier on the dataset that is highly imbalanced.

\begin{table}%[H]
	\caption{Validation results for random forests}
	\label{tab:rf_val_results}
	\centering
	\renewcommand{\arraystretch}{1}
	\begin{tabular}{lccc|ccccc}
	\hline
	&\multicolumn{3}{c}{Configuration Value} &\multicolumn{5}{c}{Cross-validation results [\%]} \\ \cline{2-9}
	&mtry&min.node.size&probability&Specificity&Recall&Precision&FPR&F1 \\
	\hline
		Conf.1 &3&1&0.5   &  99.97&	    24.36&	83.33&	0.03&	37.7 \\
		Conf.2 &6&1&0.5   &  99.96&	    25.52&	80.29&	0.04&	38.73 \\
		\rowcolor{Gray}
		\textbf{Conf.3} &\textbf{3}&\textbf{10}&\textbf{0.5}  &\textbf{99.98}&\textbf{23.67}   &\textbf{85.71}	&	\textbf{0.02}&\textbf{37.09 }\\
		Conf.4 &6&10&0.5  &  99.97&	    24.59&	83.46&	0.03&	37.99 \\
		Conf.5 &3&1&0.65  &  99.93&	    30.16&	72.63&	0.07&	42.62 \\
		Conf.6 &6&1&0.65  &  99.92&	    32.02&	70.41&	0.08&	44.02 \\
		Conf.7 &3&10&0.65 &  99.94&	    30.16&	76.47&	0.06&	43.26 \\
		Conf.8 &6&10&0.65 &  99.93&	    32.02&	72.63&	0.07&	44.44 \\
	    Conf.9 &3&1&0.8   &  99.79& 	   42&	55.35&	0.21&	47.76 \\
		Conf.10 &6&1&0.8  &  99.73& 	44.08&	50&	    0.27&	46.86	 \\
		Conf.11 &3&10&0.8  & 99.81&	    41.76&	57.32&	0.19&	48.32 \\
		Conf.12 &6&10&0.8&  99.75&	    44.32&	52.47&	0.25&	48.05	\\
		Conf.13 &3&1&0.9&    99.31&	    54.06&	32.5&	0.69&	40.59 \\
		Conf.14 &6&1&0.9&    99.19&	    54.52&	29.3&	0.81&	38.12 \\
		Conf.15 &3&10&0.9&   99.34&	    54.52&	33.76&	0.66&	41.7 \\
		Conf.16 &6&10&0.9&   99.24&	    55.22&	30.95&	0.76&	39.67 \\
	    Conf.17 &3&1&0.99&   90.67&	    83.53&	5.22&	9.33&	9.83 \\
		Conf.18 &6&1&0.99&   90.79&	    83.06&	5.26&	9.21&	9.89 \\
		\rowcolor{Gray}
		\textbf{Conf. 19} &\textbf{3}&\textbf{10}&\textbf{0.99}  &\textbf{90.31}&\textbf{84.92}   &\textbf{5.12}	&	\textbf{9.69}&\textbf{9.65 }\\
		Conf.20 &6&10&0.99&  90.63&	    83.29&	5.19&	9.37&	9.77	 \\
		\hline
	\end{tabular}     
\end{table}

\begin{table}
	\renewcommand\arraystretch{1.6}
\begin{minipage}{.45\linewidth}
	\centering
	
	\caption{Confusion matrix for Conf.\,\,3 Random forest}
	\label{tab:rf_conf1}
	
	\medskip
	\begin{tabular}{l|l|c|c|c}
		\multicolumn{2}{c}{}&\multicolumn{2}{c}{\small \bfseries Actual value}&\\
		\cline{3-4}
		\multicolumn{2}{c|}{\bfseries \small Prediction}&\small \bfseries fraud&\small \bfseries non-fraud&\multicolumn{1}{c}{ \bfseries Total}\\
		\cline{2-4}
		&\small \bfseries fraud & \small $102$ & \small $17$ & \small $119$\\
		\cline{2-4}
		& \small \bfseries non-fraud &\small  $329$ &\small $69991$ &\small $70320$\\
		\cline{2-4}
		\multicolumn{1}{c}{} & \multicolumn{1}{c}{\small \bfseries Total} & \multicolumn{1}{c}{\small $431$} & \multicolumn{    1}{c}{\small $70008$} & \multicolumn{1}{c}{\small $70439$}\\
	\end{tabular}
\end{minipage}\hfill
\begin{minipage}{.45\linewidth}
	\centering
	
	\caption{Confusion matrix for Conf.\,\,19 Random forest}
	\label{tab:rf_conf2}
	\medskip
	
	\begin{tabular}{l|l|c|c|c}
		\multicolumn{2}{c}{}&\multicolumn{2}{c}{\small \bfseries Actual value}&\\
		\cline{3-4}
		\multicolumn{2}{c|}{\bfseries \small Prediction}&\small \bfseries fraud&\small \bfseries non-fraud&\multicolumn{1}{c}{ \bfseries Total}\\
		\cline{2-4}
		&\small \bfseries fraud & \small $366$ & \small $6786$ & \small $7152$\\
		\cline{2-4}
		& \small \bfseries non-fraud &\small  $65$ &\small $63222$ &\small $63287$\\
		\cline{2-4}
		\multicolumn{1}{c}{} & \multicolumn{1}{c}{\small \bfseries Total} & \multicolumn{1}{c}{\small $431$} & \multicolumn{    1}{c}{\small $70008$} & \multicolumn{1}{c}{\small $70439$}\\
	\end{tabular}
\end{minipage}
\end{table}

\subsection{Support Vector Machine results}
For the purpose of training Support Vector Machines we chose the radial basis function as a kernel and additionally we increased cost of misclassifying samples to better address the problem of class imbalance in the dataset. The tuned parameters were: cost of constraints violation (cost) and kernel parameter gamma. As shown in Table \ref{tab:svm_val_results} SVM achieved high recall, but with quite low precision for almost all configurations. If we only consider recall, Conf 1. was better than random forests' Conf 19. with significantly higher false positive rate.  Actually, no set of parameters was able to get false positive rate lower than 10\%. If we also had to choose configuration with the lowest FPR, Conf. 20 would be the best candidate. 

%\begin{table}%[H]
%	\caption{Classification results for SVM using 10 fold cross-validation}
%	\label{tab:svm_cv_results}
%	\centering
%	\renewcommand{\arraystretch}{0.9}
%	\begin{tabular}{lrc|ccccc}
%		\hline
%		&\multicolumn{2}{c}{Configuration Value} &\multicolumn{5}{c}{Cross-validation results [\%]} \\ \cline{2-8}
%		&cost& gamma&Specificity&Recall&Precision&FPR&F1 \\
%		\hline
%		Conf 1. &	1&	0.077&	72.58&	86.95&	1.97&	27.42&	3.86\\
%		Conf 2. &	1&	0.100&	74.56&	85.84&	2.09&	25.44&	4.08\\
%		Conf 3. &	1&	0.500&	80.02&	85.35&	2.64&	19.98&	5.12\\
%		Conf 4. &	1&	1.000&	84.39&	83.40&	3.27&	15.61&	6.29\\
%		Conf 5. &	1&	2.000&	86.07&	81.56&	3.57&	13.93&	6.84\\
%		Conf 6. &	5&	0.077&	76.80&	85.26&	2.27&	23.20&	4.42\\
%		Conf 7. &	5&	0.100&	77.65&	85.01&	2.35&	22.35&	4.57\\
%		Conf 8. &	5&	0.500&	83.90&	84.77&	3.23&	16.10&	6.22\\
%		Conf 9. &	5&	1.000&	85.93&	82.53&	3.58&	14.07&	6.86\\
%		Conf 10. &	5&	2.000&	87.67&	79.94&	3.94&	12.33&	7.51\\
%		Conf 11. &	10&	0.077&	77.62&	85.14&	2.35&	22.38&	4.57\\
%		Conf 12. &	10&	0.100&	78.27&	84.78&	2.41&	21.73&	4.68\\
%		Conf 13. &	10&	0.500&	85.08&	83.92&	3.44&	14.92&	6.60\\
%		Conf 14. &	10&	1.000&	86.36&	81.93&	3.67&	13.64&	7.01\\
%		Conf 15. &	10&	2.000&	88.27&	78.94&	4.08&	11.73&	7.76\\
%		Conf 16. &	50&	0.077&	78.92&	84.96&	2.49&	21.08&	4.83\\
%		Conf 17. &	50&	0.100&	79.40&	85.11&	2.55&	20.60&	4.94\\
%		Conf 18. &	50&	0.500&	86.22&	82.49&	3.65&	13.78&	6.99\\
%		Conf 19. &	50&	1.000&	87.64&	80.73&	3.97&	12.36&	7.56\\
%		Conf 20. &	50&	2.000&	89.73&	76.85&	4.52&	10.27&	8.54\\
%		\hline
%	\end{tabular}     
%\end{table}

\begin{table}%[H]
	\caption{Validation results for SVM}
	\label{tab:svm_val_results}
	\centering
	\renewcommand{\arraystretch}{1}
	\begin{tabular}{lrc|ccccc}
	\hline
	&\multicolumn{2}{c}{Configuration Value} &\multicolumn{5}{c}{Cross-validation results [\%]} \\ \cline{2-8}
	&cost& gamma&Specificity&Recall&Precision&FPR&F1 \\
	\hline
		\rowcolor{Gray}
	\textbf{Conf 1.} &\textbf{1}&\textbf{0.077}&\textbf{72.62}  &\textbf{87.47}&\textbf{1.93}   &\textbf{27.38}	&	\textbf{3.77}\\
		Conf 2. &	1&	0.100& 75.03&	86.77&	2.09&	24.97&	4.09\\
		Conf 3. &	1&	0.500& 79.52&	84.69&	2.48&	20.48&	4.82\\
		Conf 4. &	1&	1.000& 84.38&	83.99&	3.20&	15.62&	6.17\\
		Conf 5. &	1&	2.000& 85.84&	82.60&	3.47&	14.16&	6.65\\
		Conf 6. &	5&	0.077& 76.78&	84.92&	2.20&	23.22&	4.29\\
		Conf 7. &	5&	0.100& 77.72&	84.92&	2.29&	22.28&	4.47\\
		Conf 8. &	5&	0.500& 84.00&	83.99&	3.13&	16.00&	6.04\\
		Conf 9. &	5&	1.000& 85.60&	83.76&	3.46&	14.40&	6.64\\
		Conf 10. &	5&	2.000& 87.38&	79.35&	3.72&	12.62&	7.12\\
		Conf 11. &	10&	0.077&77.64&	84.69&	2.28&	22.36&	4.44\\
		Conf 12. &	10&	0.100&78.33&	85.38&	2.37&	21.67&	4.61\\
		Conf 13. &	10&	0.500&85.07&	83.29&	3.32&	14.93&	6.39\\
		Conf 14. &	10&	1.000&85.99&	83.06&	3.52&	14.01&	6.76\\
		Conf 15. &	10&	2.000&	88.00&	76.80&	3.79&	12.00&	7.23\\
		Conf 16. &	50&	0.077&78.87&	85.15&	2.42&	21.13&	4.71\\
		Conf 17. &	50&	0.100&79.44&	85.15&	2.49&	20.56&	4.83\\
		Conf 18. &	50&	0.500&86.09&	83.06&	3.55&	13.91&	6.80\\
		Conf 19. &	50&	1.000&87.35&	80.05&	3.75&	12.65&	7.17\\
			\rowcolor{Gray}
		\textbf{Conf 20.} &\textbf{50}&\textbf{2.000}&\textbf{89.41}  &\textbf{75.87}&\textbf{4.23}   &\textbf{10.59}	&	\textbf{8.01}\\
		\hline
	\end{tabular}     
\end{table}

\subsection{XGBoost results}
\par In case of XGBoost we analyzed following hyperparameters in different configurations: maximum depth of a tree (max.depth), minimum sum of instance weight needed in a child (min.child.weight), subsample ratio of columns when constructing each tree (colsample) and, as in random forests, cut-off probability. As for the training itself, we set maximum number of iterations to 2000 with learning rate parameter set to 0.1 using early stop if error does not decrease in 100 consecutive iterations. 

Even though we built classifiers for 240 combinations of hyperparameters we decided to present only 20 most interesting. In Table \ref{tab:xgb_val_results} Conf. 1 - Conf. 10 have  the smallest false-positive rate and the other 10 configurations have significantly larger recall. Looking at the classification results we can draw a similar conclusion as in the case of random forest - cut-off probability is the most important parameter for the outcome. After examining the other parameters we were not able to clearly describe their exact impact for the results. As shown in Table \ref{tab:xgb_val_results} validation results confirmed, Conf. 1 and Conf. 16 being the best in their categories, but slightly worse than the best two random forest configurations.

\begin{table}%[H]
	\caption{Validation results for XGBoost}
	\label{tab:xgb_val_results}
	\centering
	\renewcommand{\arraystretch}{0.9}
	\begin{tabular}{lcccc|ccccc}
		\hline
		&\multicolumn{4}{c}{Configuration Value} &\multicolumn{5}{c}{Cross-validation results [\%]} \\ \cline{2-10}
		&max.depth&colsample&min.c.w&prob.&Specificity&Recall&Precision&FPR&F1 \\
		\hline
		\rowcolor{Gray}
		\textbf{Conf 1.} &\textbf{6}&\textbf{0.25}&\textbf{1}  &\textbf{0.50}&\textbf{99.95}   &\textbf{31.32}	&	\textbf{78.03}&\textbf{0.05}&\textbf{44.70}\\
		Conf 2.  &	9&	0.25&	1&	0.50&   99.95&	30.16&	78.30&	0.05&	43.55 \\
		Conf 3.  &	3&	0.25&	2&	0.50&   99.94&	31.32&	76.27&	0.06&	44.41 \\
		Conf 4.  &	3&	0.25&	1&	0.50&   99.95&	32.02&	78.41&	0.05&	45.47 \\
		Conf 5.  &	3&	0.50&	1&	0.50&   99.94&	32.71&	77.05&	0.06&	45.93 \\
		Conf 6.  &	6&	0.50&	1&	0.50&   99.94&	32.02&	76.24&	0.06&	45.10 \\
		Conf 7.  &	9&	0.50&	1&	0.50&   99.94&	32.48&	76.09&	0.06&	45.53\\
		Conf 8.  &	3&	0.75&	1&	0.50&   99.95&	33.18&	79.89&	0.05&	46.89 \\
		Conf 9.  &	6&	0.75&	1&	0.50&   99.94&	31.32&	77.14&	0.06&	44.55 \\
		Conf 10. &  9&	0.75&	1&	0.50&  99.94&	32.71&	77.05&	0.06&	45.93	 \\
		Conf 11. &	3&	0.50&	8&	0.99&  93.79&	79.35&	7.30&	6.21&	13.36 \\
		Conf 12. &	3&	0.25&	8&	0.99&  93.58&	80.51&	7.16&	6.42&	13.16	\\
		Conf 13. &	3&	0.25&	4&	0.99&  93.74&	80.05&	7.30&	6.26&	13.37 \\
		Conf 14. &	3&	0.50&	4&	0.99&  93.99&	79.58&	7.54&	6.01&	13.77 \\
		Conf 15. &	3&	1.00&	4&	0.99&  94.20&	78.42&	7.68&	5.80&	14.00 \\
		\rowcolor{Gray}
		\textbf{Conf 16.} &\textbf{3}&\textbf{1.00}&\textbf{8}  &\textbf{0.99}&\textbf{93.89}   &\textbf{80.51}	&	\textbf{7.50}&\textbf{6.12}&\textbf{13.72}\\
		Conf 17. &	3&	0.25&	1&	0.99&  93.97&	78.89&	7.45&	6.03&	13.61 \\
		Conf 18. &	3&	0.75&	8&	0.99&  93.83&	79.58&	7.35&	6.17&	13.46 \\
		Conf 19. &	3&	0.25&	2&	0.99&  93.91&	79.81&	7.47&	6.09&	13.66 \\
		Conf 20. &	3&	0.75&	4&	0.99&  94.05&	80.05&	7.65&	5.95&	13.96 \\
		\hline
	\end{tabular}     
\end{table}

\subsection{Sensitivity analysis}
Decision to conduct sensitivity analysis was motivated by our inability to indicate the exact moment of the marking any particular account as fraudulent and thus aggregated transactions data might be contaminated with transactions that happened after an alert on Etherscan has been raised for a particular account. This may lead to look-ahead bias since we are using data that was unknown at the moment of detecting a fraudulent account. In our approach we investigated what impact on the quality of the classifiers excluding the most important and potentially biased variables might have.   

Importance of considered variables is not as easily determined when using SVM as in random forest or XGBoost. Furthermore, none of the SVM results was as satisfactory (in terms of recall) as the best of random forests or XGBoost. These two observations led to omission of SVM in our sensitivity analysis.

\begin{figure}%[H]
	\includegraphics[width=\linewidth]{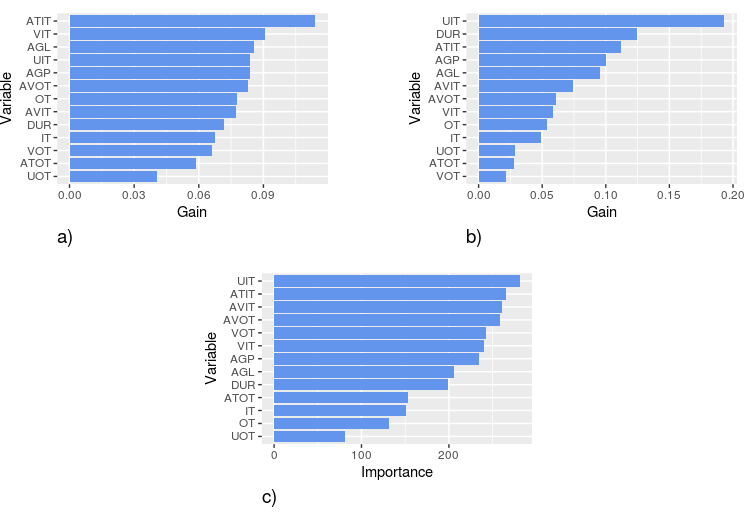}
	\caption{Variable importance for: a) XGBoost Conf. 1, b) XGBoost Conf. 16, c) Random Forest Conf. 3}
	\label{fig:importance}
\end{figure}

Explanatory variables importances were calculated separately for each of the best configurations and are presented in the Figure \ref{fig:importance}.

Considering random forests variable importance (sometimes called "gini importance") is defined as the total decrease in node impurity weighted by the probability of reaching that node averaged over all trees in the forest. Impurity is defined as:
\begin{equation}
	G = \sum_{i=1}^{C}p(i) * (1-p(i))
\end{equation}
with $C$ being the number of classes and $p(i)$ being the probability of picking a datapoint with class $i$. 

In case of XGBoost relative variable importance is measured as the Gain which is contribution of the corresponding feature to the model calculated by taking each feature's contribution for each tree in the model.
If we define $G_j = \sum_{i \in I_j} g_i$ and $H_j = \sum_{i \in I_j} h_i$ (based on the Equation \ref{eq:xgb}) where $I_j$ is the set of indices of data points assigned to the $j$-th leaf, we can express Gain as:
\begin{equation}
	Gain = \frac{1}{2}[\frac{G^2_L}{H_L + \lambda} + \frac{G^2_R}{H_R + \lambda} + \frac{(G_L + G_R)^2}{H_L + H_R + \lambda}] - \gamma
\end{equation}
This formula can be decomposed as 1) the score on the new left leaf 2) the score on the new right leaf 3) The score on the original leaf 4) regularization on the additional leaf.

As we can see in Fig. \ref{fig:importance} the most important variables for each classifier are usually connected with the incoming and the least important with the outgoing transactions. The only variable that is either first or second in terms of importance for all three classifiers is average time between incoming transactions. 
For the XGBoost we decided to apply a minor change to the chosen configurations. Instead of stopping after having no decrease of error in 100 consecutive iterations, XGBoost would do 2000 iterations regardless of the results.

\begin{table}[]
	\caption{Validation results for random forests with $n$ most important variables excluded}
	\centering
	\label{tab:rf_imp_val}
	\renewcommand{\arraystretch}{0.9}
	\begin{tabular}{lccccc}
		\hline
		&\multicolumn{5}{c}{Validation results [\%]} \\ \cline{2-6}
		&Specificity&Recall&Precision&FPR&F1 \\
		\hline
		Conf. 3 ($n$ = 2)& 99.98&	15.55&	81.71&	0.02&	26.12\\
		Conf. 3 ($n$ = 4)& 99.98&	14.62&	84&	0.02&	24.90\\
		Conf. 3 ($n$ = 8)& 99.98&	7.66&	71.74&	0.02&	13.84\\
		Conf. 19 ($n$ = 2)& 89.52&	82.37&	4.62&	10.48&	8.74\\
		Conf. 19 ($n$ = 4)& 89.38&	81.67&	4.52&	10.62&	8.57\\
		Conf. 19 ($n$ = 8)& 88.66&	68.91&	3.60&	11.34&	6.86\\
		\hline
	\end{tabular}     
\end{table}

\begin{table}%[H]
	\caption{Validation results for XGBoost with $n$ most important variables excluded }
	\label{tab:xgb_imp_val}
	\centering
	\renewcommand{\arraystretch}{0.9}
	\begin{tabular}{lccccc}
		\hline
		&\multicolumn{5}{c}{Validation results [\%]} \\ \cline{2-6}
		&Specificity&Recall&Precision&FPR&F1 \\
		\hline
		Conf. 1 ($n$ = 2)& 99.95&	26.68&	75.16&	 0.05&	39.38\\
		Conf. 1 ($n$ = 4)& 99.95&	17.63&	68.46&	 0.05&	28.04\\
		Conf. 1 ($n$ = 8)& 99.98& 	2.78&	54.55&	 0.02&	5.30  \\
		Conf. 16 ($n$ = 2)& 92.66&	76.33&	6.02&	 7.34&	11.15\\
		Conf. 16 ($n$ = 4)& 90.69&	71.46&	4.51&	 9.31&	8.49\\
		Conf. 16 ($n$ = 8)& 87.03&  62.41&	 2.88&	 12.97&	5.50  \\
		\hline
	\end{tabular}     
\end{table}

As presented in Tables \ref{tab:rf_imp_val} and \ref{tab:xgb_imp_val} random forest turned out to be more resistant to cutting off important variables. Even though false positive rate for Conf. 19 is high, with 8 variables excluded we are still able to detect almost 70\% of all frauds.
%\begin{table}
%	\caption{Comparison of chosen configurations of each models}
%	\label{tab:final_results}
%	\centering
%	\renewcommand{\arraystretch}{0.9}
%	\begin{tabular}{lccccc}
%		\hline
%		&\multicolumn{5}{c}{Validation results [\%]} \\ \cline{2-6}
%		&Specificity&Recall&Precision&FPR&F1 \\
%		\hline
%		Conf.1 random& 96.71&	67.75&	11.26&	 3.29&	19.31\\
%		Conf.2 random& 94.89&	64.73&	 7.24&	 5.11&	13.02\\
%		Conf.1 XGBoost&  99.95&	31.32&	78.03&	0.05&	44.70\\
%	Conf.11 XGBoost& 93.79&	79.35&	7.30&	6.21&	13.36 \\
%		Conf.3 RF&  99.98&	23.67&	85.71&	0.02&	37.09 \\
%		Conf.19 RF& 90.31&	84.92&	5.12&	9.69&	9.65 \\
%		Conf.1 SVM& 72.62&	87.47&	1.93&	27.38&	3.77\\
%		Conf.20 SVM& 89.41&	75.87&	4.23&	10.59&	8.01\\
%		\hline
%	\end{tabular}     
%\end{table}

\section{Conclusions and future work}
Due to the significant developments in blockchain technology, dedicated fraud prevention systems are an important area of research. We proposed a machine learning based method for predicting whether a particular account on Ethereum blockchain might be fraudulent. 

Three different classifiers were analyzed and out of them Random Forest obtained the best results in terms of recall and false positive rate separately, having the other statistics at the reasonable level (in one of the configurations SVM had the best recall for the validation set but at the same time it had three times worse false positive rate). 

Best recall for Random Forest was 84.92\%. It did not justify using this model in any real-world anti-fraud system. The reason was significant amount of type I error being made by that classifier where almost 10\% percent of all accounts would be alerted.

Configuration 3 for Random Forest that achieved 0.02\% of false positive rate was still able to detect 23.67\% of all frauds. This result can be perceived as a good candidate for an automated anti-fraud system. If we would like to deploy such a system on any cryptocurrency exchange or within cryptocurrency wallet we will mark as fraudulent one in five thousands accounts.

As for future work, we would like to obtain data from exchanges that will help determine whether proposed method can be applied in the current form or is needing further enhancements. 

Conducted sensitivity analysis showed that proposed model are not too sensitive for particular explanatory variables but one of future research directions may include estimating exact moments of marking particular account as fraudulent. Then, we would not take a risk of our training set being vulnerable to look-ahead bias.

\bibliographystyle{splncs04}
\bibliography{mybibfile}

\end{document}